\documentclass[preprint, preprintnumbers,nofootinbib,12pt]{revtex4-2}
\usepackage[english]{babel}
\usepackage{color}
\usepackage{amsmath}
\usepackage{physics}
\usepackage[hidelinks]{hyperref}
\usepackage{diagbox}
\usepackage{multirow}
\usepackage{amsfonts}
\usepackage{amssymb}
\usepackage{latexsym}
\usepackage{graphicx}
\usepackage{adjustbox}
\usepackage{xr}
\usepackage{siunitx}
\usepackage{threeparttable}
\usepackage{braket}
\usepackage[justification=centering]{caption}
\usepackage{float}
\usepackage{caption}
\usepackage{subcaption}
\usepackage{mathtools}
\usepackage{array}
\usepackage{seqsplit}
\usepackage{textcomp}
\usepackage{enumitem}
\usepackage{tikz}
\usepackage{tabularx}
\usepackage{setspace}
\usepackage{soul}
\usepackage{cancel}
\usepackage[scr=rsfs]{mathalpha}
\usepackage{amsmath,bm}

\newcommand{\la}{\lambda}

\newcommand{\rar}{\rightarrow}

\begin{document}

\title{Radial power-like potentials: from the Bohr-Sommerfeld $S$-state energies to the exact ones\\}

\author{J.C.~del~Valle}
\email{juan.delvalle@ug.edu.pl}

\affiliation{{Institute of Mathematics}, Faculty of Mathematics, Physics, and Informatics, University of Gda\'nsk, 80-308 Gda\'nsk, Poland}
\author{A.V.~Turbiner}
\email{turbiner@nucleares.unam.mx, alexander.turbiner@stonybrook.edu (corresponding author)}

\affiliation{Instituto de Ciencias Nucleares, Universidad Nacional Aut\'onoma de M\'exico,
A. Postal 70-543 C. P. 04510, Ciudad de M\'exico, M\'exico.}

\begin{abstract}
Following our previous study of the Bohr-Sommerfeld (B-S) quantization condition for
one-dimensional case (del Valle \& Turbiner (2021) \cite{First}), we extend it
to $d$-dimensional power-like radial potentials.
The B-S quantization condition for $S$-states of the $d$-dimensional radial Schr\"odinger equation is proposed.
Based on numerical results obtained for the spectra of power-like potentials, $V(r)=r^m$ with
$m \in [-1, \infty)$, the correctness of the proposed B-S quantization condition is established for various dimensions $d$. It is  demonstrated that by introducing the {\it WKB correction} $\gamma$ (supposedly coming from the higher order WKB terms) into the r.h.s. of the B-S quantization condition leads to the so-called {\it exact WKB quantization condition}, which reproduces the exact energies, while $\gamma$ remains always very small. For $m=2$ (any integer $d$) and for $m=-1$ (at $d=2$) the WKB correction $\gamma=0$: for $S$ states the B-S spectra coincides with the exact ones.

Concrete calculations for physically important cases of linear, cubic, quartic, and sextic
oscillators, as well as Coulomb and logarithmic potentials in dimensions $d=2,3,6$ are presented. Radial quartic anharmonic oscillator is considered briefly.

\end{abstract}
\maketitle
\newpage
\section*{Introduction}

In our previous paper \cite{First}\textcolor{blue}{,} a detailed analysis of the Bohr-Sommerfeld (B-S) quantization condition for one-dimensional power-like potentials $V(x)=|x|^m, m>0$ was carried out. The absolute/relative deviations of the B-S energies from the exact ones for ground state and (highly)-excited states were calculated. It was demonstrated that even for the low-lying excited states these energies are reasonably close not only for highly-excited states. Maximal deviation occurs for the ground state.
It was also shown that a simple modification of the r.h.s. of the B-S quantization condition for confining potentials $V(x)$,
\begin{equation}
	\frac{1}{\hbar}\int_{x_A}^{x_B}\sqrt{E-V(x)}\,dx\ =\ \pi\left(N\ +\ \frac{1}{2} + \gamma(N) \right)\ ,\qquad N\ =\  0,1,2, ...\ ,
	\label{1D}
\end{equation}
where $x_{A,B}$ are turning points and $N$ is quantum number, by introducing a small, dimensionless, numerically calculable function $\gamma=\gamma(N)$, which we called the {\it WKB correction}, leads to exact energy spectra. It was checked that for various power-like potentials $|\gamma| \leq 1/2$\,. Evidently, for the harmonic oscillator $\gamma=0$, while for the square well potential with infinite wells $\gamma=1/2$ and for highly-excited states it decays, $\gamma(N) \sim 1/N$ as $N \rar \infty$. The goal of the present paper is to generalize this observation and extend it to the $S$-states of the $d$-dimensional confining power-like radial potentials.

Needless to say that from the early days of quantum mechanics there were numerous attempts to generalize the B-S quantization condition (\ref{1D}) to the case of $d$-dimensional radial potentials.
In this work, we propose the generalized B-S quantization condition
\begin{equation}
	\frac{1}{\hbar}\int_{0}^{r_0}\sqrt{E_{BS}-V(r)}\,dr\ =\
     \pi\left(n_r\ +\ \frac{d}{4}\right)\ ,\qquad n_r \ =\  0,1,2, ...\ .
\label{radial}
\end{equation}
for $S$-(bound) states, where $r_0$ is the radial turning point. Here $n_r$ is radial quantum number, it is equal to the number of radial nodes in the eigenfunction of the $d$-dimensional radial Schr\"odinger equation, namely\footnote{Boundary conditions are chosen in such a way that
$\int_{0}^{\infty}|\psi(r)|^2r^{d-1}\,dr\,<\, \infty\ , \quad d=1,2,...\ $ and $|\psi(r)| \neq \infty $ at
$r \rar 0$.}
\begin{equation}
	 	-\frac{\hbar^2}{2\mu}\left(\psi''(r)\ +\ \frac{d-1}{r}\psi'(r)\right)+\ V(r)\psi(r)\ =\
        E\psi(r)\ , \qquad r\in[0,\infty) 	\ ,
\label{radial_se}
\end{equation}
for $S$-states, where for the sake of simplicity we put $\mu=1/2$. Note for $d=2$, the quantization condition (\ref{radial}) was derived in \cite{Berry,Brack} using the Langer transformation.  For arbitrary $d$, a quantization condition\footnote{Same l.h.s. as in (\ref{radial}), but with an $m$-dependent r.h.s.} similar to (\ref{radial}) was proposed in the article \cite{Marinov}, see Appendix therein. However,
it was established only for the singular potentials with behavior $V(r) \sim r^{m}$ at $r \rar 0$,
when $-2 < m < 0$. Following these works \cite{Berry,Brack,Marinov} we propose the quantization condition (\ref{radial}) for the general confining, not necessarily singular at origin potentials.

There are three fundamental properties behind the proposed quantization condition (\ref{radial}):
\begin{enumerate}
	\item At $d=1$, it coincides with the celebrated one-dimensional WKB quantization
     condition\footnote{With $N=2\,n_r$.} assuming an even potential, $V(x)=V(-x)$ and $x_A=-x_B$,
	\item For the harmonic oscillator potential, $V(r)=r^2$, the condition (\ref{radial}) leads
    to the exact energies for any $d$,
    \item At large $n_r$\textcolor{blue}{,} the Bohr-Sommerfeld energies $E_{BS}$ from (\ref{radial}) approach to the exact ones.
\end{enumerate}	

Naturally, the WKB correction $\gamma$, see (\ref{1D}), can be introduced into r.h.s. of (\ref{radial}),
\begin{equation}
	\frac{1}{\hbar}\int_{0}^{r_0}\sqrt{E-V(r)}\,dr\ =\ \pi\left(n_r\ +\ \frac{d}{4}\ +\
       \gamma\right)\ .
\label{radial_modified}
\end{equation}
Namely, this formula will be the main object of the study in the present paper. We will call this formula the {\it exact WKB (radial) quantization condition}.

For the sake of simplicity the general power-like radial potential of the form\footnote{Here $a$ is a dimensionless  parameter and $g$ a coupling constant. }
\begin{equation}
		V(r)\ =\ a\,g^{m-2}\,\frac{m}{|m|}\,r^m\ ,\ m \geq -1\ ,
\label{rpot}
\end{equation}
will be studied. For those potentials with $m>0$, the integral in the l.h.s. of (\ref{radial}) and (\ref{radial_modified}) is reduced to the Euler Beta function and the energies can be written in closed analytic form. The case $m \leq 0$ requires special attention and it will be discuss in Section \ref{SectionII} for two specific cases.  We will follow the same program as in our previous study \cite{First}: as the first step by making comparison the energies emerging from the B-S quantization condition (\ref{radial}) with ones obtained from highly-accurate numerical calculations in the equation (\ref{radial_se}), and we calculate the absolute and relative deviations the B-S and exact energies. As the second step, we calculate the WKB correction $\gamma$ in (\ref{radial_modified}) for different power-like potentials in different dimensions, which allows us to reproduce the exact energies.

The structure of the paper is the following.  Section \ref{SectionI} is devoted to studying the analytical and numerical properties of $E_{BS}$ for the power-like potentials at $m>0$. The explicit calculations of the WKB correction for  the physically relevant dimensions, $d=2,3,6$ are presented.  In Section \ref{SectionII}, we apply the quantization rule (4) to the singular potentials with $m\leq0$: the Coulomb and logarithmic potentials.  We explore the accuracy of $E_{BS}$ and compute the WKB correction.  The results are summarized in Conclusions.

\section{Non-Singular Power-Like Potentials}
\label{SectionI}

In this Section, without loss of generality we set $a=g=1$ in (\ref{rpot}), and consider vanishing at origin radial potentials
\begin{equation}
		V(r)\ =\ r^m\ ,\ m > 0\ \textcolor{blue}{,}
\label{rpot-positive_m}
\end{equation}
where $r=\sqrt{x_1^2+x_2^2+\ldots+x_d^2}$ is the radius in $R^d$. Interestingly,
the potential (\ref{rpot-positive_m}) coincides with the Green function for $(2-m)$-dimensional gravity, being the potential produced by point-like mass. The quantization condition  (\ref{radial}) for potential (\ref{rpot-positive_m})
can be solved analytically, it leads to
\begin{equation}
    E_{BS}^{(m,d)}\ =\ \left( 2\hbar MB\left(\tfrac{1}{2},M\right)\left(n_r+\frac{d}{4}\right)\right)^{\frac{1}{M}}\ ,\qquad M\ =\ \frac{1}{m}\ + \ \frac{1}{2}\ ,
\label{BSradial}
\end{equation}
where $B(a,b)$ is the Euler Beta function, see \cite{NIST}. In what follows, we set $\hbar=1$. Note that the expression for $E_{BS}$ may be analytically continued from integer $n_r$ to complex $n_r$.
By construction, $E_{BS}$ (\ref{BSradial}) leads to the exact energy spectrum \cite{Exact}
for the radial harmonic oscillator $m=2$:
\begin{equation}
  E_{BS}^{(m=2,d)} \ =\  4n_r+d\ =\ E_{exact}^{(m=2,d)}\ .
\label{E-m=2}
\end{equation}

In the limit $m \rar \infty$, the  potential (\ref{rpot-positive_m}) becomes a $d$-dimensional radial square-well potential with an infinite wall at $r=1$. In this limit, the Bohr-Sommerfeld energies are given by
\begin{equation}
   E_{BS}^{(m=\infty,d)}\ =\ \pi^2\left(n_r\ +\ \frac{d}{4}\right)^2\ .
\label{BSWell}
\end{equation}
Let us note that at $d=1$,
\begin{equation}
	E_{exact}^{(m=\infty,d=1)}\ =\ \pi^2\left(n_r+\frac{1}{2}\right)^2\ ,
\end{equation}
which coincides with the expression emerging directly from (1).
For $d>1$, the exact energies obey the equation
\begin{equation}
J_{\frac{d-2}{2}}\left(\sqrt{E_{exact}^{(m=\infty,d>1)}}\right)\ =\ 0\  ,
\label{Bessel}
\end{equation}
where $J_\nu$ is the Bessel function of the first kind of order $\nu$, see \cite{NIST}.
At $d=3$ the roots of the Bessel function $J_{\frac{1}{2}}$ are known
in closed analytic form\footnote{In fact, $J_{\frac{1}{2}}(x)\ \propto\ \frac{1}{\sqrt{x}}\sin x$.} and the energies read
\begin{equation}
\ E_{exact}^{(m=\infty,d=3)}\ =\ \pi^2(n_r+1)^2\ .
	\end{equation}
It is worth pointing out that by replacing $d/4$ in the r.h.s. of (\ref{BSWell}) by $(d+1)/4$\,, $E_{BS}^{(m=\infty,d)}$ coincides with $E_{exact}^{(m=\infty,d)}$ at $d=1,3$.
For $d \neq 1\ \mbox{or}\ 3$, Eq.(\ref{Bessel}) can be solved by numerical means only. For arbitrary $d$,
and large $n_r$,
\begin{equation}
     E_{exact}^{(m=\infty,d)}\ =\ \pi^2\left(n_r+\frac{d+1}{4}\right)^2\ +\ \mathcal{O}(n_r^{-1})
\label{exact}
\end{equation}
in accordance with the asymptotic expansion for the zeros of $J_{\frac{d-2}{2}}$, see \cite{NIST}.

For fixed $m$ and large $d$, it can be checked that the leading term of the exact energy in $1/d$-expansion reads,
\begin{equation}
	E_{exact}^{(m,d=\infty)}\ =\ c_M\,d^{\frac{1}{M}}\ +\ \ldots \ \ ,
\end{equation}
where
\begin{equation}
     c_M\ =\  2^{1-\frac{1}{M}} \left(\frac{1}{2 M-1}\right)^{1-\frac{1}{2 M}}M\ .
\label{coeff1}
\end{equation}
In turn, $E_{BS}$ (\ref{BSradial}) leads to
\begin{equation}
 \qquad c_M^{(BS)}\ =\
 \left(\frac{1}{2}M B(M-\tfrac{1}{2},M)\right)^{\tfrac{1}{M}}\ .
\label{coeff2}
\end{equation}
In general, the coefficients (\ref{coeff1}) and (\ref{coeff2}) are different. However, in the  limit $m\rar0$ they coincide, $c_M=c_M^{(BS)}=1$.

\subsection{Numerical Results}

Now, we can compare the Bohr-Sommerfeld energies $E_{BS}$ (\ref{BSradial}) with the exact ones. As the examples it is done explicitly for linear, cubic, quartic, and sextic oscillators ($m=1,3,4,6$, respectively) in dimensions  $d=2$ and $d=3$, see Tables \ref{d=2} and \ref{d=3}, respectively. Four quantities are shown: the exact energy $E_{exact}^{(m,d)}$, the Bohr-Sommerfeld energies $E_{BS}^{(m,d)}$, and the absolute \text{(A.D.)} and relative \text{(R.D.)} deviations,
\begin{equation}
\label{A+R.D}
	\text{A.D.}\ =\ |E_{exact}-E_{\text{BS}}|\quad ,
     \quad \text{R.D.}\ =\ \frac{\left|E_{exact}-E_{\text{BS}}\right|}{E_{exact}}\ ,
\end{equation}
respectively. Note that the exact energies $E_{exact}^{(m,d)}$ for quantum numbers $n_r=0,...,40$ were calculated using the Lagrange Mesh Method \cite{Baye,delValle}. Based on the computational code developed in our previous study \footnote{Let us note that this code can be easily adapted to tackle the two and three-dimensional radial Schr\"odinger equations (\ref{radial_se}), see \cite{delValle}.} \cite{First}, we carried out calculations with {  100, 500, 1000 mesh points}. For all studied potentials (\ref{rpot-positive_m}) it was checked explicitly that 100 (optimized) mesh points is sufficient to obtain  {  four exact decimal digits (d.d.) (and sometimes more digits) for all studied eigenstates with $n_r=(0 - 40)$ at $d=1,2,3,6$. In general, the increase of the number of mesh points leads to dramatic increase in the accuracy in the energies, see below, for discussion see \cite{Second} and the book \cite{QAHO}}.

\begin{figure}[h]
    	\includegraphics[]{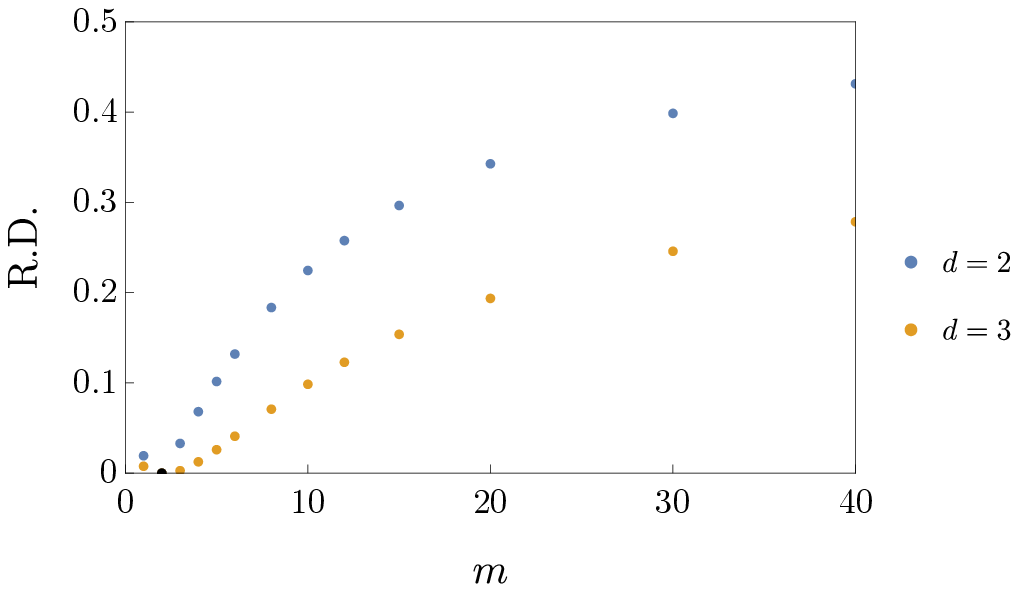}
    	\caption{R.D. of the Bohr-Sommerfeld ground state ($n_r=0$) energy  from the exact one for potential $V(r) =r^m$ {\it vs.} $m$ for $d=2,3$. Points correspond to $m = 1, 2, 3, 4, 5,6, 8, 10, 12, 15, 20, 30, 40$. Minimal deviation reached at $m=2$. Maximal deviations,  0.577 $(d=2)$ and 7/16 ($d=3$), reached at $m\rar\infty$ .}
\label{fig:RD}
\end{figure}

According to Tables \ref{d=2} and \ref{d=3} the exact energies $E_{exact}$ are always larger than the B-S energies $E_{BS}$ for $m=3,4,6$ at $d=2,3$. Of course, they coincide, $E_{exact}=E_{BS}$,
at $m=2$. It hints that the similar behavior should be observed for larger integer $m > 6$. Evidently, these results can be extended to non-integer $m > 2$. At $m=1$, a distinction between $d=2$ and $d=3$ cases occurs: the $E_{BS}$ energies approach to exact ones from above for $d=2$, while  for $d=3$ they approach from below.  Similar behavior seems to occur for all potentials for $0 < m < 2$.  For a given $n_r$,  relative accuracy increases for $m > 2$. For the ground state, $n_r=0$,  this can be seen in Fig. \ref{fig:RD} for $d=2,3$, where R.D. is presented as the function of $m$. For $m = 2$ we have R.D.=0, and then, R.D. begins to grow: for $m < 2$, when $m$ is decreasing, and for $m > 2$, when $m$ is increasing.

\begin{table}[h]
	\caption{Linear, cubic, quartic, and sextic oscillators  at $d=2$. For $E_{exact}$ and $E_{BS}$ all printed digits are exact,  no rounding. }
	{\setlength{\tabcolsep}{0.4cm}			
		\begin{tabular}{ccccccc}
			\hline	
			\rule{0pt}{4ex}				
			$m$&$n_r$  & $E_{exact}$ & $E_{BS}$&A.D.& R.D&$\gamma$  \\[5pt]
			\hline	
			\multirow{4}{*}{1}
			\rule{0pt}{4ex}	
			& 0&1.7372 &1.7706&$3.3\times10^{-2}$&$1.9\times10^{-2}$&$-1.4\times10^{-2}$  \\
			& 5&8.7545 &8.7579&$3.4\times10^{-3}$&$3.8\times10^{-4}$&$-3.2\times10^{-3}$  \\
			& 10&13.4761 &13.4778&$1.6\times10^{-3}$&$1.2\times10^{-4}$&$-1.9\times10^{-3}$  \\
			& 20&21.0530 &21.0537&$7.7\times10^{-4}$&$3.4\times10^{-5}$&$-1.1\times10^{-3}$   \\[10pt]
			\multirow{4}{*}{3}
			& 0&2.1874 &2.1154&$7.2\times10^{-2}$&$3.3\times10^{-2}$&$1.4\times10^{-2}$  \\
			&5  &37.6011  &37.5896&$1.2\times10^{-2}$&$3.1\times10^{-4}$&$1.4\times10^{-3}$  \\
			&10 &81.6763 &81.6695&$6.8\times10^{-3}$&$8.3\times10^{-5}$&$7.2\times10^{-4}$  \\
			&20  &182.2834 &182.2794&$4.0\times10^{-3}$&$2.2\times10^{-5}$&$3.7\times10^{-4}$  \\[10pt]
			\multirow{4}{*}{4}
			&0  &2.3448  &2.1850&$1.6\times10^{-1}$&$6.8\times10^{-2}$&$2.7\times10^{-2}$  \\
			&5  &53.4863  &53.4550&$3.1\times10^{-2}$&$5.9\times10^{-4}$&$2.4\times10^{-3}$  \\
			&10 &126.6175  &126.5972&$2.0\times10^{-2}$&$1.6\times10^{-4}$&$1.3\times10^{-3}$  \\
			&20  &308.9313  &308.9183&$1.3\times10^{-2}$&$4.2\times10^{-5}$&$6.5\times10^{-4}$  \\[10pt]
			\multirow{4}{*}{6} &
			0  &2.6093  &2.2650 &$3.4\times10^{-1}$&$1.3\times10^{-1}$&$4.9\times10^{-2}$ \\
			&5  &82.7310  &82.6369&$9.4\times10^{-2}$&$1.1\times10^{-3}$&$4.2\times10^{-3}$  \\
			&10 &218.0469  &217.9788&$6.8\times10^{-2}$&$3.1\times10^{-4}$&$2.2\times10^{-3}$  \\
			&20  &594.6983  &594.6495&$4.9\times10^{-2}$&$8.2\times10^{-5}$&$1.1\times10^{-3}$  \\[10pt]
			\hline	
	\end{tabular}}
	\label{d=2}
\end{table}

\begin{table}[h]
	\caption{Linear, cubic, quartic, and sextic oscillators  at $d=3$. For $E_{exact}$ and
             $E_{BS}$ all printed digits are exact,  no rounding. }
	{\setlength{\tabcolsep}{0.4cm}			
\begin{tabular}{ccccccc}
			\hline	
			\rule{0pt}{4ex}				
			$m$&$n_r$  & $E_{exact}$ & $E_{BS}$&A.D.& R.D&$\gamma$   \\[5pt]
			\hline	
			\multirow{4}{*}{1}
			\rule{0pt}{4ex}	
			& 0&2.3381 &2.3202&$1.8\times10^{-2}$&$7.6\times10^{-3}$&$8.7\times10^{-3}$  \\
			& 5&9.0226 &9.0213&$1.3\times10^{-3}$&$1.4\times10^{-4}$&$1.2\times10^{-3}$  \\
			& 10&13.6914 &13.6909&$5.6\times10^{-4}$&$4.1\times10^{-5}$&$6.5\times10^{-4}$ \\
			& 20&21.2248 &21.2245&$2.3\times10^{-4}$&$1.1\times10^{-5}$&$3.4\times10^{-4}$ \\[10pt]
			\multirow{4}{*}{3}			
			&0  &3.4505  &3.4411&$9.4\times10^{-3}$&$2.7\times10^{-3}$&$1.7\times10^{-3}$  \\
			&5  &39.6535  &39.6492&$4.3\times10^{-3}$&$1.1\times10^{-4}$&$5.3\times10^{-4}$  \\
			&10 &84.0111  &84.0084&$2.7\times10^{-3}$&$3.2\times10^{-5}$&$2.8\times10^{-4}$  \\
			&20 &184.9517 &184.9501&$1.6\times10^{-3}$&$8.7\times10^{-6}$&$1.5\times10^{-4}$ \\[10pt]
			\multirow{4}{*}{4}
			&0  &   3.7996  &3.7519&$4.8\times10^{-2}$&$1.3\times10^{-2}$&$7.1\times10^{-3}$  \\
			&5  &  56.7342  &56.7190&$1.5\times10^{-2}$&$2.7\times10^{-4}$&$1.1\times10^{-3}$  \\
			&10 & 130.6420  &130.6320&$1.0\times10^{-2}$&$7.7\times10^{-5}$&$6.2\times10^{-4}$ \\
			&20 & 313.9580 &313.9515 &$6.6\times10^{-3}$&$2.1\times10^{-5}$&$3.2\times10^{-4}$ \\[10pt]
			\multirow{4}{*}{6}
			&0  &   4.3385&4.1612&$1.8\times10^{-1}$&$4.1\times10^{-2}$&$2.1\times10^{-2}$    \\
			&5  &  88.3923  &88.3348&$5.8\times10^{-2}$&$6.5\times10^{-4}$&$2.5\times10^{-3}$  \\
			&10 & 225.8520 &225.8099&$4.2\times10^{-2}$&$1.9\times10^{-4}$&$1.3\times10^{-3}$  \\
			&20 & 605.5907  &605.5604&$3.0\times10^{-2}$&$5.0\times10^{-5}$&$6.9\times10^{-4}$ \\[10pt]
			\hline	
\end{tabular}}
\label{d=3}
\end{table}

\subsection{WKB Correction}

Analogously to the one-dimensional case, we introduce the WKB correction $\gamma=\gamma(n_r,d)$ through
\begin{equation}
	\frac{1}{\hbar}\int_{0}^{r_0}\sqrt{E_{exact}-V(r)}\,dr\ =\
              \pi\left(n_r\ +\ \frac{d}{4}\ +\ \gamma(n_r,d) \right)\ ,
\label{radial_modified-1}
\end{equation}
cf.(\ref{radial_modified}), where $\gamma$ is defined in such a way that (\ref{radial_modified-1}) becomes the exact WKB quantization condition;
the turning points are located at $r=0, r_0$.

\begin{figure}[h]
    	\includegraphics[]{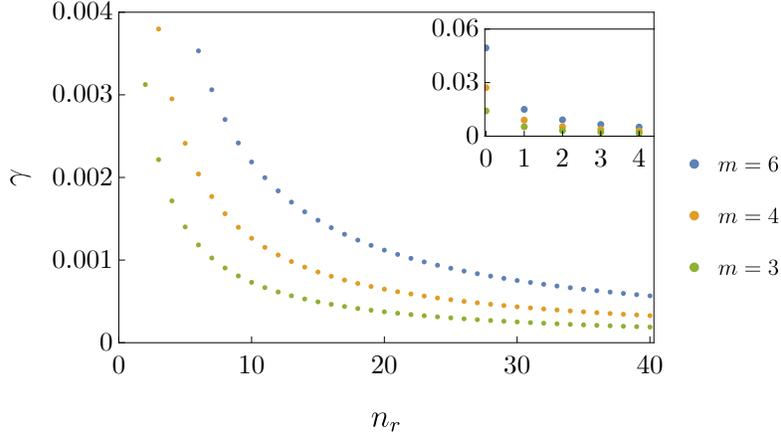}
    	\caption{WKB correction $\gamma$ {\it vs} $n_r$ for cubic $m=3$, quartic $m=4$, and sextic $m = 6$ two-dimensional $(d=2)$ potentials, see (12). Displayed dots correspond to $n_r$ = 0,1,2,\ldots,40. In subFigure the domain $n_r = 0, 1, 2, 3, 4$ shown.}
\label{fig:radial_gamma_d2}
\end{figure}

\begin{figure}[h]
    	\includegraphics[]{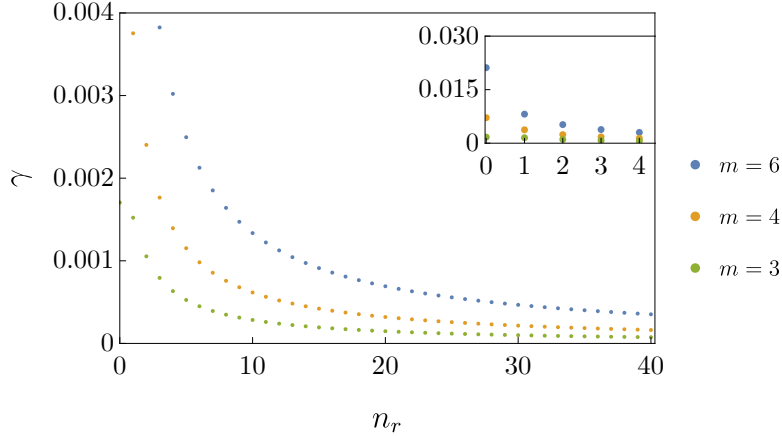}
    	\caption{WKB correction $\gamma$ {\it vs.} $n_r$ for cubic $m=3$, quartic $m=4$, and sextic $m = 6$ three-dimensional $(d=3)$ potentials, see (12). Displayed dots correspond to $n_r$ = 0,1,2,\ldots,40. In subFigure the domain $n_r = 0,1,2,3,4$ shown.}
\label{fig:radial_gamma_d3}
\end{figure}

For power-like potentials $V=r^m$ (\ref{rpot-positive_m}) with $m>0$, the integral in l.h.s. of (\ref{radial_modified-1}) can be evaluated analytically: it leads to the $\gamma$-modified B-S energies
\begin{equation}
	E_{BS}^{(modified)}\ =\ \left(2\hbar
       MB\left(\tfrac{1}{2},M\right)\left(n_r+\frac{d}{4}+\gamma \right)\right)^{\frac{1}{M}}\ ,
        \quad M\ =\ \frac{1}{m}\ + \ \frac{1}{2}\ ,
\label{BS}
\end{equation}
cf.(\ref{BSradial}). The WKB correction, $\gamma=\gamma(n_r,m,d)$ is determined in the same way as for the one-dimensional case: by making a comparison between the accurate $E_{exact}$ at fixed $m$, $n_r$, $d$, and $E_{BS}^{(modified)}$ with requirement that
\[
   E_{BS}^{(modified)}\ =\ E_{exact}\ .
\]
It allows to find $\gamma$ unambiguously. This procedure was realized for the linear, cubic, quartic and sextic radial oscillators in dimensions $d=2,3$.
For these potentials the results for $\gamma$ are shown in Tables \ref{d=2} and \ref{d=3}. The plots of the WKB correction $\gamma$ {\it vs.} $n_r$ are shown in Figs.\ref{fig:radial_gamma_d2} and \ref{fig:radial_gamma_d3} for $m=3,4,6$.  {  In general, the energies defined
with 4-5 significant digits allow us to obtain $\gamma$ with accuracy of 4-5 significant digits.
Concrete calculations of the spectra for linear, cubic, quartic and sextic oscillators at dimension $d=2,3,6$ in Lagrange Mesh Method with 100 (optimized) mesh points \cite{Second} lead to $\gamma$ with 4-5 correct significant digits. }
All curves are very small, smoothly decreasing with growth of $n_r$ functions with maximum at $n_r=0$. Note that for a given $m$, the $\gamma$ is about ten times smaller in comparison with one from the one-dimensional case \cite{First}. According to Tables \ref{d=2} and \ref{d=3}, $\gamma$ seems to decrease as $d$ grows for given $m$ and $n_r$.

\begin{figure}[h]
	\includegraphics[]{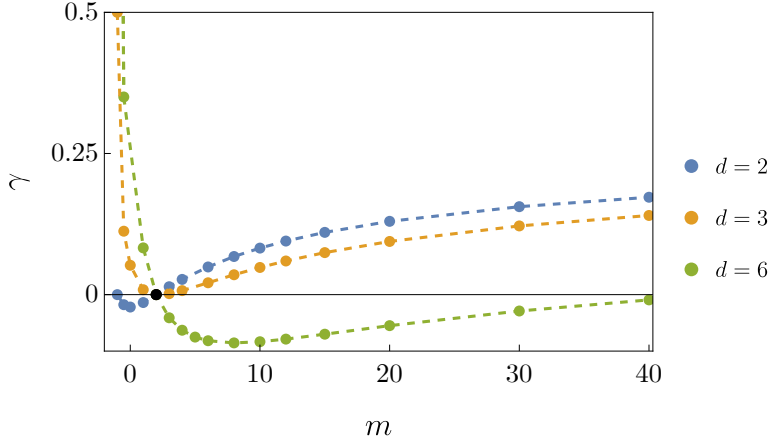}
\caption{WKB correction $\gamma$ {\it vs.} $m$, see (\ref{BS}), for the ground state $n_r = 0$ and $d=2,3,6$. Displayed dots correspond to $m =-1,-1/2,0 (\log r-potential), 1,2,3,4,5,6,8,10,12,15,20,30,40$.
The black bullet at $m=2$ corresponds to $\gamma=0$ at any $d$.}
\label{fig:Gammavsm}
\end{figure}

Note that for fixed $m$ and $d$ all WKB corrections studied show a very smooth behavior as functions of the quantum number $n_r$. Moreover, they can be easily fitted using the similar functions, which {  were} employed in one-dimensional case, see \cite{First}.  For example, for the quartic radial oscillator $m=4$ at $d=3$, the excellent fit is achieved with
\begin{equation}
	\gamma_{fit}^{(m=4)}(n_r,d=3)\ =\ \frac{0.014361n_r\ +\ 0.00344}{\sqrt{n_r^4\ +
                    \ 1.99078 n_r^3\ +\ 1.31424 n_r^2\ +\ 0.45314 n_r\ +\ 0.02645}}\ .
\label{fit4}
\end{equation}
If substituted into the exact quantization condition (\ref{BS}), (\ref{fit4}) leads to the energies with, at least, $5$ exact d.d. for $0\leq n_r \leq 100$. In general, {
the accuracy in the energy and in $\gamma$ are related: a higher accuracy in the energy leads
to higher accuracy in $\gamma$ and vice versa. In particular, for the radial oscillator with arbitrary $m > 0$ and integer $d$, in order to reach a higher accuracy in the energies, the $\gamma$ should be fitted with a higher accuracy by using the function
\begin{equation}
\gamma_{fit}\ =\ \frac{P_k(n_r)}{\sqrt{Q_{2k+2}(n_r)}}\ ,
\label{fit}
\end{equation}
with $k>1$, where $P_k(n_r), Q_{2k+2}(n_r)$ are polynomials of degrees $k$ and $2k+2$, respectively.	
}
For $n_r=0$ the behavior of $\gamma$ {\it vs.} $m$ is shown on Fig.\ref{fig:Gammavsm} for $d=2,3,6$. Note that for $d=6$,  $\gamma(m, n_r=0)=0$ occurs at $m=2$ and $m \approx 45.8$, while for $d=2$, $\gamma(m, n_r=0)=0$ occurs at $m=2$ and $m=-1$, see next Section.

\section{Singular Potentials}
\label{SectionII}

\subsection{Coulomb Potential}

Let us consider the potential (\ref{rpot}) at $m=-1$, it corresponds to the Coulomb potential
\begin{equation}
	V(r)\ =\ -\frac{1}{r}\ .	
	\label{CoulombPot}
\end{equation}
In this case the radial equation (\ref{radial_se}) can be solved exactly, its eigenvalues are
\begin{equation}
\label{m=-1 - exact}
  E_{exact}\ = -\frac{1}{\left(2n_r + d-1\right)^2}\ ,
\end{equation}
for $S$ states.
The B-S quantization condition (\ref{radial}) can also be solved explicitly resulting in
\begin{equation}
\label{m=-1 - BS}
	E_{BS}\ =\ -\frac{1}{\left(2n_r+\dfrac{d}{2}\right)^2}\ . 	
\end{equation}
By making a comparison of (\ref{m=-1 - exact}) and (\ref{m=-1 - BS}),
one can immediately conclude that the WKB correction
\begin{equation}
	\gamma\ =\ \frac{d}{2} - 1\ ,
\end{equation}
and $\gamma > 0$ for $d>2$. For any $d > 2$ and fixed $n_r$ the inequality $E_{BS} < E_{exact}$ holds.

It is worth emphasizing that for $d=2$  the WKB correction $\gamma=0$, hence, for $S$-states the Bohr-Sommerfeld spectra $E_{BS}$ in $two$-dimensional space coincides with the exact spectra $E_{exact}$ of two-dimensional Coulomb problem. This observation was in missed in \cite{Berry,Brack}. It can be used for
construction of the analogue of the B-S quantization condition (\ref{radial}) for the states with $L \neq 0$ for $d=2$.

\subsection{Logarithmic Potential}

The logarithmic potential\footnote{It can be formally regarded as the power-like potential at $m=0$ by noting that
		$\log r\ =\ \lim\limits_{m\rar0}\left(\frac{1}{m}r^m\ -\ \frac{1}{m}\right)$.}
\begin{equation}
V(r)\ =\ \log r\ ,
\label{log}
\end{equation}

For the potential (\ref{log}), the quantization condition (\ref{radial}) can be solved exactly, leading  to
\begin{equation}
E_{BS}\ =\ \log\left[2\sqrt{\pi}\left(n_r\ +\ \frac{d}{4}\right)\right]\ .
\label{EBS_Log}
\end{equation}
In Table \ref{Logarithmic}, a comparison between $E_{exact}$ and $E_{BS}$ for $d=2,3,6$ is shown
together with results for A.D., R.D. and $\gamma$. {  All printed digits for $E_{exact}$ are exact, they are found with 100 (optimized) mesh points in the Lagrange Mesh method and checked with 500 mesh points calculation.}
Contrary to what occurred for power-like potentials at $m \neq 0$, the A.D. and R.D. increase with growth of $d$. On the other hand, $\gamma$ is a slowly decreasing, smooth function as $n_r$ increases. It can be interpolated by
\begin{equation}
\label{fit0-d2}
\gamma_{fit}^{(\log r)}(n_r,d=2)=\frac{-0.00327\,n_r-0.02243}{\sqrt{0.00015n_r^4+0.74891n_r^2+1.15229n_r+1}}\ ,
\end{equation}
for $d=2$ and
\begin{equation}
\label{fit0-d3}
  \gamma_{fit}^{(\log r)}(n_r,d=3)\ =\
  \frac{0.00636\,n_r + 0.05157}{\sqrt{0.00002\,n_r^4 + 0.11072\,n_r^2 + 0.98319\,n_r+1}}\ ,
\end{equation}
for $d=3$,
\begin{equation}
\label{fit0-d6}
{
	\gamma_{fit}^{\log r}(n_r,d=6)\ =\ \frac{0.012166 n_r+0.283345}{\sqrt{0.000002 n_r^4+0.016873 n_r^2+0.411624 n_r+1}}
}
\end{equation}
{
for $d=6$, cf. (\ref{fit4}), with accuracy 4 d.d.} Note that in the denominators in
(\ref{fit0-d2}),(\ref{fit0-d3}),{  (\ref{fit0-d6}) the coefficients in front of $n_r^3$ vanish and other coefficients decrease with increase of $d$ if constant is kept equal to 1.}
If these fits are inserted into the exact quantization condition (\ref{radial_modified-1}), this leads to the energies in (\ref{radial_modified-1}) with not less than 4 exact s.d.
Accuracy in the fit of $\gamma$ can be gradually increased by using the fitting function (\ref{fit}) with $k = 2,3,4 \ldots$.
Final expression for the energies
\begin{equation}
E_{exact}\ =\ \log\left[2\sqrt{\pi}\left(n_r\ +\ \frac{d}{4}\ +\ \gamma \right)\right]\ .
\label{E_exact_Log}
\end{equation}

\begin{table}[h]
	\caption{Logarithmic potential   at $d=2,3,6$. For $E_{exact}$ and $E_{BS}$ all printed digits are exact,  no rounding. }
	{\setlength{\tabcolsep}{0.4cm}			
		\begin{tabular}{ccccccc}
			\hline	
			\rule{0pt}{4ex}				
			$d$&$n_r$  & $E_{exact}$ & $E_{BS}$&A.D.& R.D&$\gamma$   \\[5pt]
			\hline	
			\multirow{4}{*}{2}
			\rule{0pt}{4ex}	
			& 0&0.5265 &0.5724&$4.6\times10^{-2}$&$8.7\times10^{-2}$&$-2.2\times10^{-2}$  \\
			& 5&2.9688&2.9702&$1.4\times10^{-3}$&$4.7\times10^{-4}$&$-8.0\times10^{-3}$  \\
			& 10&3.6163 &3.6168&$5.0\times10^{-4}$&$1.4\times10^{-4}$&$-6.2\times10^{-3}$  \\
			& 20&4.2857 &4.2859&$2.0\times10^{-4}$&$4.7\times10^{-4}$&$-4.9\times10^{-3}$   \\[10pt]
			\multirow{4}{*}{3}			
			&0 &1.0443  &0.9778&$6.7\times10^{-2}$&$6.4\times10^{-2}$&$5.2\times10^{-2}$  \\
			&5  &3.0196  &3.0147&$4.9\times10^{-3}$&$1.6\times10^{-3}$&$2.8\times10^{-2}$   \\
			&10 &3.6427  &3.6404&$2.3\times10^{-3}$&$6.3\times10^{-4}$&$2.5\times10^{-2}$  \\
			&20  &4.2990 &4.2980&$1.0\times10^{-3}$&$2.3\times10^{-4}$&$2.0\times10^{-2}$  \\[10pt]
						\multirow{4}{*}{6}			
			&0 &1.8443  &1.6709&$1.7\times10^{-1}$&$9.4\times10^{-2}$&$2.8\times10^{-1}$  \\
			&5  &3.1653  &3.1373&$2.8\times10^{-2}$&$8.8\times10^{-3}$&$1.8\times10^{-1}$  \\
			&10 &3.7212  &3.7078&$1.3\times10^{-2}$&$3.6\times10^{-3}$&$1.6\times10^{-1}$ \\
			&20  &4.3396 &4.3335&$6.0\times10^{-3}$&$1.4\times10^{-3}$&$1.3\times10^{-1}$  \\[10pt]
			\hline	
	\end{tabular}}
	\label{Logarithmic}
\end{table}
\section*{Conclusions}

A new quantization rule for the $d$-dimensional radial potentials is proposed for the case of $S$-states. The two basic requirements behind this quantization condition are accomplished:
(i) the coincidence with the one-dimensional standard WKB quantization condition at $d=1$;
(ii) the spectrum of the $d$-dimensional harmonic oscillator should be exact. Explicit calculations for power-like potentials, including  Coulomb and logarithmic potentials, prove the correctness of the proposal. For the Coulomb potential at $d=2$ the obtained BS energies coincide with exact ones. {  In order to obtain the eigenvalues accurately, the radial Schr\"odinger equation was solved using the Lagrange Mesh method with 100, 500, 1000 mesh points. It allows easily to reach extremely high accuracies, e.g. for linear potential with 100 (optimized) mesh points, the energies of the first 37 low-lying $S$ states can be established with at least 4 exact d.d. while with 500 optimized mesh points, the energies of the first 100 low-lying $S$ states can be established with at least 140 exact d.d.}

The notion of the WKB correction, introduced for $d=1$ case, was extended to the $d$-dimensional radial case. For power-like radial potentials, $V(r)=r^m$ with fixed $m$ and quantum number $n_r$, the WKB correction gets smaller as $d$ increases.
For fixed $d$ and $m$, the WKB correction is a monotonous decreasing function of $n_r$.
For fixed $d$ and $n_r$, the WKB correction increases as $m>2$ grows. In general, the WKB correction $\gamma$ is a smooth, dimensionless function of parameters $(d,m,n_r)$: it can an easily fitted by ratio of a polynomial to square-root of another polynomial of the argument $n_r$. In all studied cases $\gamma < 1/2$.

It is interesting to extend the analysis from power-like potentials to two-term anharmonic oscillators
$V=r^2 + \la r^m$, where $\la$ is the coupling constant, which are widely used in applications.
It will be done elsewhere. The simplest non-trivial case of the quartic anharmonic oscillator, see Table \ref{anharmonic}, shows that for the ground state energies the WKB correction $\gamma$ remains (very) small for all $\la \geq 0$ at $d=1,2,3,6$.

\begin{table}
                \caption{WKB correction for the ground state energy in $V(r)=r^2 + \la r^4$
                at $d=1,2,3,6$ for different coupling constants $\la$. Exact ground state energies
                from the book \cite{QAHO}. Results rounded to the
                first-second significant digits. }
        {\setlength{\tabcolsep}{0.4cm}
\begin{tabular}{|c|cccccc|}
\hline
\hline
                \rule{0pt}{4ex}
                $\la$ & 0 & $0.1$ & $1$ & $10$ & $100$&$\infty$ \\[5pt]
                \hline\rule{0pt}{4ex}
                $\gamma(\la,d=1)$ & 0 &  0.014  &  0.049  &  0.072  &  0.079  & 0.081
\\[5pt]
                $\gamma(\la,d=2)$ & 0 &  0.008  &  0.020  &  0.025  &  0.027  & 0.027
\\[5pt]
                $\gamma(\la,d=3)$ & 0 &  0.005  &  0.008  &  0.007  &  0.007  & 0.007
\\[5pt]
                $\gamma(\la,d=6)$ & 0 & -0.021  & -0.047  & -0.059  & -0.062  &-0.063
\\[5pt]
\hline
\hline
\end{tabular}}
\label{anharmonic}
\end{table}	

Present authors are not aware how to calculate $\gamma$ analytically from the first principles.

\section*{Acknowledgments}

\noindent
Partial financial support to J.C.dV. was provided by the SONATABIS-10 grant \\
no.2019/34/E/ST1/00390 (Poland). A.V.T. was partially supported by CONACyT grant A1-S-17364 and
DGAPA grant IN113022 (Mexico). A.V.T. thanks B Eynard, G Korchemsky, A Voros for useful discussions.

\bibliography{references.bib}

\end{document}